\documentclass[aps,prb,reprint,twocolumn]{revtex4}
\usepackage{amsmath}    
\usepackage{graphicx}   
\usepackage{verbatim}   
\usepackage{color}      
\usepackage{subfigure}  
\usepackage{hyperref}   
\usepackage{units}
\usepackage{upgreek}
\usepackage{color}

\begin{document}
\title{Charge transport across metal/molecular (alkyl) monolayer-Si junctions
is dominated by the LUMO level}
\author{Omer Yaffe}
\affiliation{Dept. of Materials \& Interfaces, Weizmann Inst. of Science, Rehovot 76100, Israel}
\author{Yabing Qi}
\affiliation{Dept. of Electrical Engineering, Princeton University, Princeton, New Jersey, 08544 USA}
\author{Lior Segev}
\affiliation{Dept. of Materials \& Interfaces, Weizmann Inst. of Science, Rehovot 76100, Israel}
\author{Luc Scheres}
\affiliation{Lab. of Organic Chemistry, Wageningen Univ., Dreijenplein 8, 6703 HB Wageningen, The Netherlands}
\author{Sreenivasa Reddy Puniredd}
\affiliation{Department of Chemical Engineering, Technion - Israel Institute of Technology, Haifa 32000, Israel}
\author{Tal Ely}
\affiliation{Dept. of Materials \& Interfaces, Weizmann Inst. of Science, Rehovot 76100, Israel}
\author{Hossam Haick}
\affiliation{Department of Chemical Engineering, Technion - Israel Institute of Technology, Haifa 32000, Israel}
\author{Han Zuilhof}
\affiliation{Lab. of Organic Chemistry, Wageningen Univ., Dreijenplein 8, 6703 HB Wageningen, The Netherlands}
\author{Leeor Kronik}
\affiliation{Dept. of Materials \& Interfaces, Weizmann Inst. of Science, Rehovot 76100, Israel}
\author{Antoine Kahn}
\affiliation{Dept. of Electrical Engineering, Princeton University, Princeton, New Jersey, 08544 USA}
\author{Ayelet Vilan}
\affiliation{Dept. of Materials \& Interfaces, Weizmann Inst. of Science, Rehovot 76100, Israel}
\author{David Cahen}
\affiliation{Dept. of Materials \& Interfaces, Weizmann Inst. of Science, Rehovot 76100, Israel}

\date{\today}

\begin{abstract}
We compare the charge transport characteristics of heavy doped p$^{++}$- and n$^{++}$-Si-alkyl chain/Hg junctions. Photoelectron spectroscopy (UPS, IPES and XPS) results for the molecule-Si band alignment at equilibrium show the Fermi level to LUMO energy difference to be much smaller than the corresponding Fermi level to HOMO one. This result supports the conclusion we reach, based on negative differential resistance in an analogous semiconductor-inorganic insulator/metal junction, that for both p$^{++}$- and n$^{++}$-type junctions the energy difference between the Fermi level and LUMO, i.e., electron tunneling, controls charge transport. The Fermi level-LUMO energy difference, experimentally determined by IPES, agrees with the non-resonant tunneling barrier height deduced from the exponential length-attenuation of the current.
\end{abstract}

\maketitle
\section{Introduction}

Molecular electronics describes charge transport processes whereby molecules serve as active elements (e.g., rectifiers, switches, sensors) or passive ones (resistors or surface passivating agents) in electronic devices \cite{heath_molecular_2009}. Since the emergence of this field, \cite{mann_tunneling_1971,aviram_molecular_1974} it has expanded greatly and now includes several types of device configuration (e.g., two-terminal junctions\cite{vilan_soft_2002,venkataraman_single-molecule_2006,chen_electron_2009,tal_electron-vibration_2008}, three-terminal junctions \cite{yu_moleculeelectrode_2003} and electrochemical devices \cite{ciampi_functionalization_2007}), where the molecules serve as direct current carriers \cite{ho_choi_electrical_2008} or indirectly affect the electrical properties of a junction \cite{haick_discontinuous_2004,vilan_moleculemetal_2003}.

Despite the diversity of investigated systems, fundamental questions regarding \emph{the mechanisms for electrical current passing through molecules between two electrodes and the possibility of gaining predictive power and control over the electrical properties of molecular junctions} remain mainly unsolved.  Detailed discussions on the limitations of existing transport models can be found elsewhere.\cite{cui_changes_2002,engelkes_length-dependent_2004,huisman_interpretation_2009, paulsson_thermoelectric_2003,vilan_analyzing_2007}  In short, even for one of the simplest systems, i.e. that of alkyl chains between Au electrodes, there is a large discrepancy between the average tunnel barrier height, extracted from current - voltage ($I-V$) measurements ($\sim \unit[1.2]{eV}$), \cite{akkerman_electrical_2008,akkerman_electron_2007,cui_making_2002,engelkes_length-dependent_2004,wang_mechanism_2003}  and the barrier that is expected from the experimentally determined electrode work function and alkyl monolayer ionization potential, i.e. the energy difference between the Fermi level and highest occupied molecular orbital (HOMO) as found by Ultraviolet Photoelectron Spectroscopy, UPS ($\sim \unit[5]{eV}$).\cite{alloway_interface_2003} In spite of this apparent discrepancy, HOMO-dominated transport is the prevailing concept for such junctions. \cite{cui_changes_2002,engelkes_length-dependent_2004,huisman_interpretation_2009,paulsson_thermoelectric_2003,li_charge_2008,beebe_measuring_2008} In the present study we challenge this concept and find that transport is dominated by the energy difference between the Fermi level and the lowest unoccupied molecular orbital (LUMO) of the molecules.

It would seem that having a semiconductor (SC) instead of a metal as one of the electrodes in a molecular junction increases the difficulty in understanding the mechanisms that govern charge transport through the junction. Indeed, with moderately doped SCs, the SC depletion layer has a large effect on the overall observed current and if the band bending is very large, it completely overwhelms any molecular effect.\cite{yaffe_molecular_2009} As we shall see, though, the intrinsic asymmetry of a junction with a SC, with respect to charge carrier type, provides unique information on the energy levels involved in transport, \emph{in ways that are impossible when using only metal electrodes}.

To find how charge is transported through a monolayer of saturated alkyl chains we measure and analyze electrical transport across heavy-doped semiconductor - molecular insulator / metal junctions. Our results provide unequivocal evidence that charge transport in these junctions is controlled by the energy difference between the metal Fermi level and the LUMO of the molecule, $|E_{\mathrm{F}}-\mathrm{LUMO}|$, rather than by the energy difference with the HOMO, $|E_{\mathrm{F}}-\mathrm{HOMO}|$. This finding agrees with complementary spectroscopic measurements of the samples without Hg contact, but challenges the prevailing concept of HOMO-dominated transport. \cite{cui_changes_2002,engelkes_length-dependent_2004,huisman_interpretation_2009,paulsson_thermoelectric_2003,li_charge_2008,beebe_measuring_2008}
Highly doped n$^{++}$- and p$^{++}$- Si-C$_{n}$H$_{2n+1}$/Hg junctions ($n=2,14,16,18$) were prepared by alkylation of freshly etched P- or B-doped ($N_{\mathrm{d/a}}\approx\unit[10^{19}]{cm^{-3}}$) Si$(1\,1\,1)$ surfaces. Detailed descriptions of the preparation and characterization of both 'long' (C14-C18) and 'short' alkyl chain (C2) samples can be found elsewhere, and are given in the Suppl. Inf. (section 1 and section 2). \cite{puniredd_highly_2008,yaffe_hg/molecular_2010}

\section{Results and discussion}

\begin{figure}
  \includegraphics{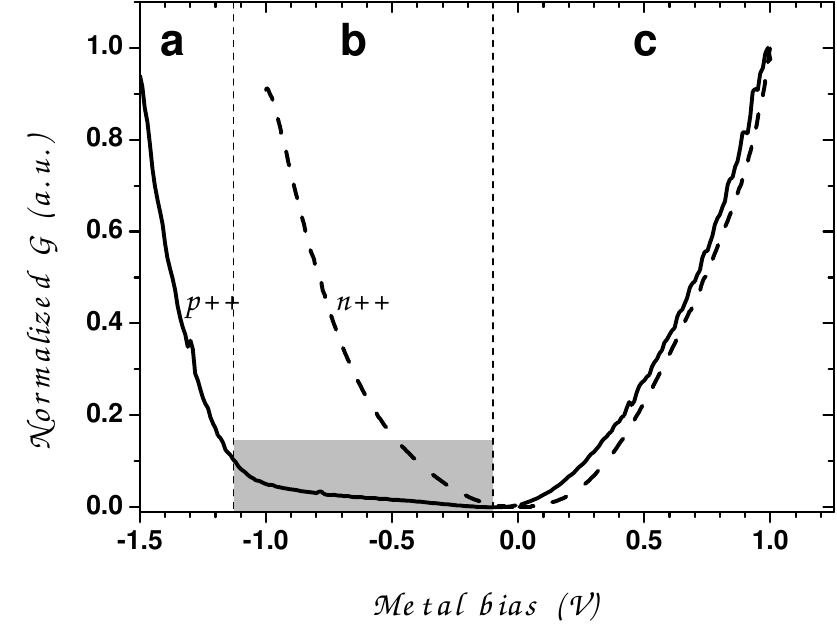}\\
  \caption{Normalized G-V curves for p$^{++}$- (solid line) and n$^{++}$- (dashed line) Si-C$_{16}$H$_{33}$/Hg junctions. The conductance of the n$^{++}$-junction is parabolic around $\unit[0]{V}$, while for the p$^{++}$-junction at negative bias on the metal there is a voltage range, where the conductance increases slowly (shaded area), followed by a sharp increase. The transition bias between the above-mentioned regions is $\sim \unit[1.1]{V}$, very close to the Si band gap ($\sim \unit[1.12]{eV}$). (a), (b) \& (c) refer to voltage ranges, shown in Fig.\ 2 and discussed in the text, for the p$^{++}$ junction only.}\label{NormCond}
\end{figure}

In Fig.\ \ref{NormCond}, the normalized conductance of p$^{++}$-/n$^{++}$-Si-C$_{16}$H$_{33}$/Hg junctions is compared on a linear scale.  While conductance for the n$^{++}$ junction (dash) is parabolic around $\unit[0]{V}$ (i.e. similar to the behaviour expected for Metal -Insulator -Metal (MIM) tunnel junctions \cite{brinkman_tunneling_1970}, for the p$^{++}$ junction (solid) there is a negative bias range where conductance increases slowly (shaded), followed by a sharp increase at $\sim \unit[1.1]{V}$, which closely corresponds to the Si band gap ($\unit[1.12]{eV}$) \cite{sze_physics_2007}.

\begin{figure}
  \includegraphics{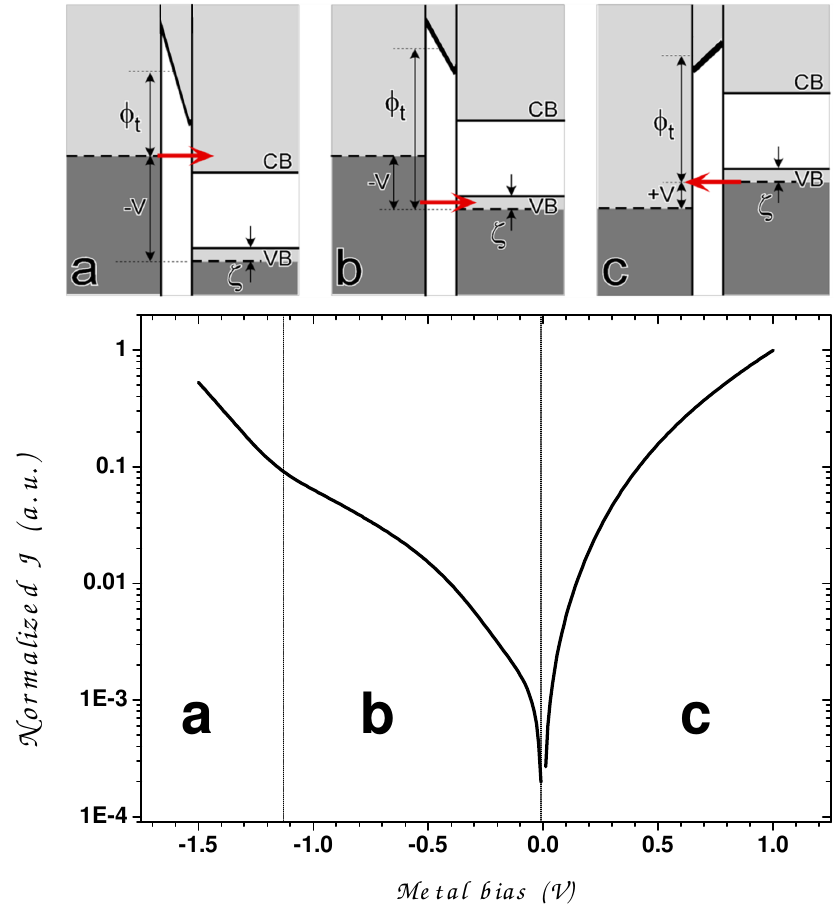}
  \caption{Schematic band energy diagrams for MIS structures with  p$^{++}$ semiconductor substrates and LUMO-dominated transport \emph{(Top)} and semi-log J-V curve of a typical p$^{++}$Si-C$_{16}$H$_{33}$/Hg junction (averaged over 12 scans) \emph{(Bottom)} with the (a), (b) and (c) labels corresponding to the diagrams above. Three major bias regimes are considered in the different energy schemes (top) and separated in the $J-V$ plot by vertical dotted lines \emph{(bottom)}, corresponding to a) metal Fermi above (i.e., closer to vacuum level) the SC gap ($V<-(E_{\mathrm{g}}+\zeta$)); b) metal Fermi within the SC gap ($-(E_{\mathrm{g}}+\zeta)<V<-\zeta$); and c) metal Fermi below the SC gap ($V>-\zeta$ ), where $E_{\mathrm{g}}$ and $\zeta$ are the forbidden gap and the energy difference between the SC Fermi level and the valence band edge. TOP: $\phi_t$  is the energy difference between the tunneling electrons (marked by a horizontal arrow) and the molecular level (spatially averaged), which is taken as the estimated effective tunneling barrier height. The insulator level is assumed to vary across the insulator. Electron tunneling into the VB (as in b) is practically identical to holes tunneling in the opposite direction. Surface states, band-bending and image forces are neglected for simplicity.}\label{Esaki}
\end{figure}

\begin{figure}
  \includegraphics{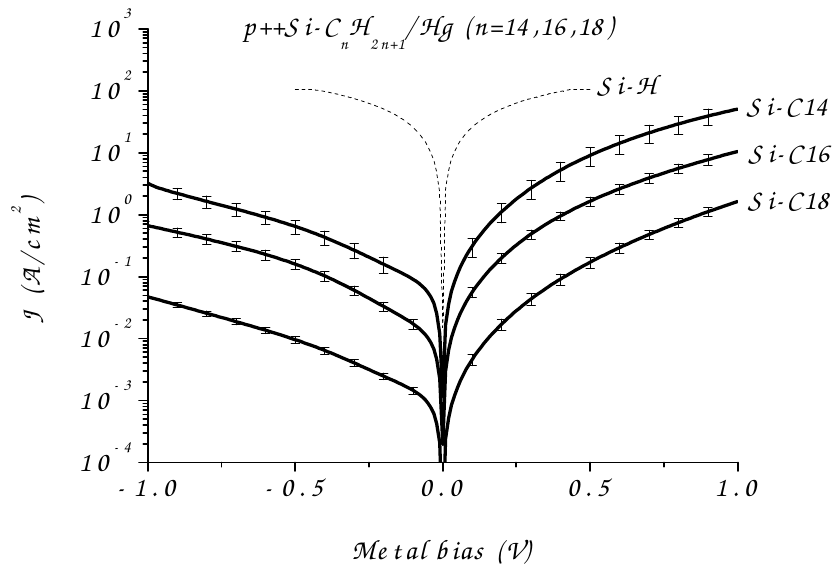}\\
  \caption{Length dependence of current density vs. voltage curves for for p$^{++}$ Si-C$_{n}$H$_{2n+1}$/Hg junctions ($n=14,16,$ and 18). Data for the Si-H/Hg junctions (dashed) are given as reference. The current density exponentially decreases with the molecular length. Error bars represent the standard deviation of at least 8 different junctions.  }\label{JVlong}
\end{figure}

This behaviour is highly reproducible and independent of the molecular length (C$_{14}$H$_{29}$-C$_{18}$H$_{37}$; see ref.\ \onlinecite{yaffe_hg/molecular_2010}  for $J-V$ curves of n$^{++}$-Si data and Fig.\ \ref{JVlong} below for the p$^{++}$-Si ones). The results of the p$^{++}$ junction at negative applied bias can be understood by considering work on Metal-inorganic Insulator- (near) degenerate Semiconductor (MIS) tunnel diodes by Esaki, Sze, Dahlke and co workers,\cite{dahlke_tunneling_1967,esaki_long_1973,esaki_new_1966,freeman_theory_1970}  which forms the basis for the present study. Schematic band diagrams of the relevant model are shown in Fig.\ \ref{Esaki} (top), along with an experimentally measured semi log $J-V$ curve of a typical p$^{++}$Si-C$_{16}$H$_{33}$/Hg junction. We identify the undulating behaviour at low to moderate negative bias (region b of Fig.\ \ref{Esaki}, which is equivalent to the shaded region in Fig.\ \ref{NormCond}) with a region where negative differential resistance (NDR) was predicted originally for the MIS system studied by Esaki et al. That effect is smeared out in the $J-V$ curve, because of the effect of interface states, an effect that, together with band bending and image forces, is, for simplicity's sake, neglected in the schematic band diagrams. In the figure, only $E_{\mathrm{F}}$ -LUMO (i.e., electron tunneling) dominated transport is considered (LUMO is depicted by thick, black line) and other cases are discussed below.

As long as the metal Fermi level does not align with the SC forbidden energy gap (i.e., regions a, c) transport is completely equivalent to tunneling in the more common Metal-Insulator-Metal (MIM) tunnel junction. \cite{simmons_generalized_1963,stratton_volt-current_1962-1}  In  regions a and c, the bias acts to reduce the averaged tunneling barrier by $\sim |V| /2$. With a moderate negative voltage to the metal (Fig.\ \ref{Esaki}) the metal Fermi level moves to energies that are in the forbidden band gap of the SC. For such a case, neglecting interface states-assisted tunneling (ISAT), Esaki and Stiles predicted an effective increase in the tunnel barrier height $\phi_{\mathrm{t}}$ (i.e., $\phi_{\mathrm{t}}$ (b) $>\phi_{\mathrm{t}}$ (a,c)), and NDR, as has indeed been observed experimentally.\cite{esaki_new_1966}  If ISAT is not neglected, instead of NDR a $\ln (J) -V$ plot will show undulating behaviour (as we show in Fig.\ \ref{Esaki}, voltage range (b), for molecular junctions), which is strongly influenced by the interface state density and distribution in the band gap. \cite{freeman_theory_1969} 
In our case ISAT has a clear influence on the $J-V$ data and, therefore, we observe undulating $J-V$ behaviour, instead of pure NDR.
Undulating $J-V$ occurs because in region b the highest energy electrons have no matching states to tunnel into, and transport proceeds via deeper electrons, aligned with the SC VB. For those electrons the barrier is now increasing as $\sim|V|/2$  instead of decreasing with $|V|$, as in band-to-band tunneling (ranges a,c). In contrast to p$^{++}$ junctions, n$^{++}$ junctions are not predicted to show undulating $J-V$ plots for LUMO-dominated tunneling because the SC Fermi and the tunneling levels are on the same side of the forbidden SC gap, regardless of applied bias (see Fig.\ S3.1b of Suppl. Inf.). Simply put, undulating $J-V$ is predicted to occur if tunneling proceeds via carriers of opposite type to the SC majority carriers.  This prediction is confirmed experimentally in Fig.\ \ref{NormCond}, which shows a highly symmetrical $G-V$ curve for the n$^{++}$-Si junction.

If the $|E_{\mathrm{F}}-\mathrm{HOMO}|$ energy difference did determine the transport barrier (i.e., hole, rather than electron tunneling), then the special undulating $J-V$ behaviour should be observed in the n$^{++}$ junction for positive applied bias on the metal and the p$^{++}$ junction should not present such behaviour at all.  The reason for the latter is that for a p$^{++}$-SC - based junction and hole tunneling,
\begin{itemize}
  \item {with negative bias applied to the metal, holes tunnel from the SC valence band to the metal;}
  \item {with positive bias applied to the metal, holes tunnel from the metal to the SC valence band.}
\end{itemize}
Neither of these flows involves tunneling of holes into the SC band gap. Thus, the undulating $\ln (J)-V$ behaviour (Fig.\ 2) and the approximately constant conductance (Fig.\ \ref{NormCond}) for the p$^{++}$-Si-alkyl/Hg junction and lack of it in the n$^{++}$ one (Fig.\ \ref{NormCond}) provide direct evidence for $|E_{\mathrm{F}}-\mathrm{LUMO}|$ (or electron)-, rather than $|E_{\mathrm{F}}-\mathrm{HOMO}|$ (hole)-dominated transport in both doping types. To put this in other words, if transport were HOMO-dominated, then the tunneling level and the SC Fermi level would be below and above the n$^{++}$-SC gap (at positive bias), while both the HOMO and the SC Fermi level would be always below the SC gap for p$^{++}$ junctions. Thus, tunneling via HOMO predicts undulating $J-V$ for n$^{++}$ junctions, which is not observed experimentally (cf. Fig.\ in the Suppl. Inf. for energy schemes that illustrate these alternatives). We note that, for an MIM junction, this type of asymmetry cannot be observed, because it originates in variation in available density of states for tunneling.
Indeed, Scott et al have used this exact asymmetry and demonstrated that inelastic tunneling is much more pronounced in a  metal/molecule-p$^{++}$Si junction than in a metal-molecule-metal junction \cite{janes_gold/molecule/p_2010}     

\begin{figure*}
  \includegraphics [scale=0.9]{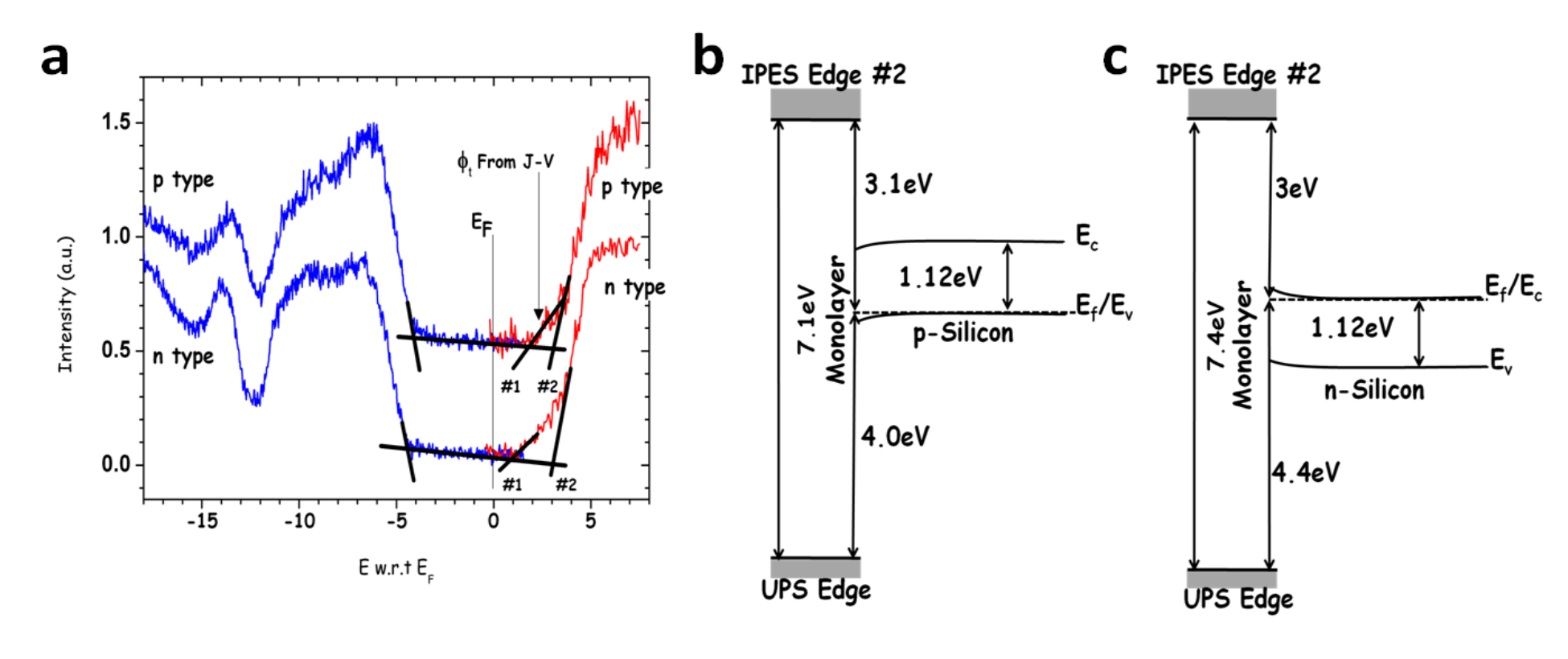}
  \caption{a) Combined UPS (left of $E_{\mathrm{F}}$ line, blue) and IPES (right to $E_{\mathrm{F}}$ line, red) spectra of C$_{16}$H$_{33}$-Si$(1\,1\,1)$ surfaces of heavy-doped p$^{++}$- (upper) and n$^{++}$- (lower) Si. The vertical black line marks the experimentally determined Fermi level of both samples. The crossings of the tangent lines determine the edges of the monolayer edge-to-edge gap. While the UPS measurements yield a single unambiguous edge, which corresponds with the position of the theoretically calculated HOMO,\cite{segev_electronic_2006}the IPES spectra show two transitions, marking two possible band edges, which are noted as \#1 and \#2, for both p$^{++}$ and n$^{++}$. The black vertical arrow marks the position of the estimated tunneling barrier relative to the Fermi level, extracted from length dependent $J-V$ measurements of the p$^{++}$Si-alkyl/Hg junction. 
b) and c) Proposed band diagrams of p$^{++}$- and n$^{++}$-Si- C$_{16}$H$_33$ surfaces, respectively, which are extracted from the edges of the UPS and IPES results. The Si band lines (solid) are slightly bent to reflect the experimentally determined $\unit[0.2]{eV}$ band bending in both samples and the dashed line marks the postion of Fermi level in the Si band gap. The grey shaded areas mark the estimated uncertainties for the positions of the actual HOSO and LUSO (see text).
  }\label{UPSIPES}
\end{figure*}

The conclusion that transport is dominated by the $|E_{\mathrm{F}}-\mathrm{LUMO}|$ energy difference implies that the LUMO is closer in energy to the SC $E_{\mathrm{F}}$ for both doping types. This can be verified by characterizing the energetics of the molecularly modified Si, prior to metal contact deposition. To this end we use \emph{Ultra-violet Photoemission Spectroscopy} (UPS), \emph{Inverse Photoemission Spectroscopy }(IPES), \emph{X-ray Photoelectron Spectroscopy }(XPS) and \emph{Contact potential difference} (CPD) measurements. 
We then study the Si band bending at equilibrium after top metal deposition by measuring charge transport across molecular monolayers (Si-C$_{2}$H$_{5}$/Hg) that are so thin that they do not present an effective tunneling barrier. Finally, we perform a molecular monolayer length-dependent study of SiC$_{n}$C$_{2n+1}$/Hg junctions where n=14, 16 and 18. These monolayers are thick enough to make tunneling the dominant current transport process. From this length dependence study we extract the tunneling barrier height and compare it to the barrier that is expected from the UPS/IPES results.

We begin with determining the band bending (BB) in Si in the p$^{++}$- and n$^{++}$-Si-C$_{16}$H$_{33}$ systems. The position of the Fermi level is known in our system from frequently repeated UPS and IPES measurements of the Fermi step on atomically clean metal (e.g. Au) surfaces. Given the nearly $\unit[20]{\AA}$  thickness of the SAM layer, the Si valence band maximum (VBM) and conduction band minimum (CBM) are not visible in UPS and IPES, and we must use here the XPS measurement of the position of the Si 2p core level, which is clearly detectable through the SAM layer, to evaluate the VBM position and the BB. Indeed, the energy difference between the core level and the VBM, called here  (Si2p-VBM) is fixed, and we can use the position of either one to determine that of the other. Combined UPS/XPS measurements done in our laboratory on carefully prepared hydrogen-terminated Si surfaces give  (Si2p-VBM)$=\unit[98.9]{eV}$, and we use this value here for the determination of the BB. Note that, in spite of being obtained with different experimental systems on differently prepared surfaces, this value compares well with that published in the literature.\cite{himpsel_determination_1983} The Si 2p peaks are found at $\unit[99.1]{eV}$ and $\unit[99.8]{eV}$ below the Fermi level for the p$^{++}$- and n$^{++}$-Si- C$_{16}$H$_{33}$ surfaces, respectively. Thus, simple considerations give:
\begin{eqnarray*}
\mathrm{BB_{p-Si}} &=& (E_\mathrm{F} - \mathrm{VBM}) \\
&=& \unit[99.1]{eV} - \unit[98.8]{eV} = \unit[0.2]{eV} \\
\mathrm{BB_{n-Si}} &=& E_{\mathrm{gap}} - (E_\mathrm{F} - \mathrm{VBM}) \\
&=& \unit[1.1]{eV} - (\unit[99.8]{eV} - \unit[98.9]{eV}) = \unit[0.2]{eV}
\end{eqnarray*}

Both p-$^{++}$ and n$^{++}$-Si surfaces are therefore seen to be slightly depleted with BB$=\unit[0.2]{eV}$. The small BB on both types of substrate is consistent with the known ability of these alkyl monolayers to passivate Si.\cite{yaffe_hg/molecular_2010} The CBM in each case is obtained by adding the $\unit[1.1]{eV}$ Si band gap to the position of the VBM.

From the position of the photoemission onset of the He(I)($\unit[21.22]{eV}$) UPS spectra, we find that the work functions of p$^{++}$- and n$^{++}$-Si-C$_{16}$H$_{33}$ are $\unit[(4.7\pm0.1)]{eV}$ and $\unit[(4.2\pm0.1)]{eV}$, respectively. Similarly, the CPD measurements performed in the dark also yield a $\unit[0.5]{eV}$ difference between the two surfaces, in excellent agreement with the photoemission data. Incidentally, CPD also shows these values to be independent of the number of carbons in the alkyl chain (2,14,16,18) for a given substrate (i.e.  p$^{++}$- or n$^{++}$-Si).The  difference between  work functions of p$^{++}$- and n$^{++}$-Si-C$_{16}$H$_{33}$ ($\unit[0.5]{eV}$) is smaller than that expected from the difference in the Si 2p core level positions for the two surfaces ($\unit[0.7]{eV}$). This may indicates that the Si-C bond polarization, which contributes to the interface dipole, is affected by the position of the Si Fermi level. Strictly speaking, the HOMO and LUMO are concepts adapted from a single molecule gas phase model, while here we use them for a monolayer in close contact with a substrate. Therefore, in the remaining of the text, we refer to them as HOSO and LUSO, highest occupied and lowest unoccupied system orbital, \cite{nesher_effect_2007} similar to HOMO* and LUMO* in ref.\ \onlinecite{engelkes_length-dependent_2004}.

Fig.\ \ref{UPSIPES} presents the combined results of He(II)($\unit[40.8]{eV}$) UPS and IPES measurmenst both for p$^{++}$ (top) and n$^{++}$-Si-C$_{16}$H$_{33}$. We define the edges of the monolayer band gap as the crossing between the tangent lines of the spectra. While the edge of the UPS spectrum is defined clearly (both for p$^{++}$ and n$^{++}$), from the IPES spectra, two different edges can be defined. The reason is that the IPES incident electrons have a larger penetration depth than the escape depth of the UPS electrons. As a result, the interface states are more pronounced in the IPES spectrum than in the UPS spectrum. These states, which result from hybridization between the alkyl molecular states and the Si VB and CB states at the interface are known as Induced Density of Interface States (IDIS) and were calculated to be localized in the first three carbons of the alkyl chain (with respect to the Si surface).\cite{segev_electronic_2006}  IDIS is probably the main physical origin for the onset of the intensity in the IPES spectrum (Fig.\ \ref{UPSIPES}). \footnote{The computational results show that the IDIS/LUMO states ratio is much larger than the IDIS/HOMO states} As a result, the IPES spectrum has a ``two slope'' behavior, noted as \#1 and \#2 in Fig.\ \ref{UPSIPES} for both the p$^{++}$ and n$^{++}$ samples. We view slope \#1 as expressing the IDIS and slope \#2 originating from the actual LUSO.   

We note that experimentally determined edges of the gap do not necessarily correspond exactly to the HOSO and LUSO of the monolayer. The onsets of UPS and IPES intensity, which define the gap edges, can be affected by spectral broadening, due to polarization of the molecules in the dense monolayer,
\cite{hill_charge-separation_2000,sato_polarization_1981,wu_electron-hole_1997}  by IDIS, \cite{segev_electronic_2006} and by radiation damage.\cite{oliver_seitz_doping_2008,amy_radiation_2006} The HOSO and LUSO positions can be found by fitting the calculated density of states results, as reported by Segev et al. for the Si-C$_{n}$H$_{2n+1}$ system \cite{segev_electronic_2006} to the experiment. In their calculations Segev et al., identified the actual LUSO to be $\sim\unit[0.5]{eV}$ above the edge, defined by slope \#2 . The actual HOSO was identified
to be $\sim\unit[0.3]{eV}$ below the UPS edge. The main implication of this calculation for the present study is that, while there is some experimental uncertainty regarding the actual position of the HOSO, the LUSO and band gap width, this uncertainty is unidirectional (i.e., the experimental edge to edge gap is the lower limit of the actual gap) and a similar situation holds for both UPS (HOSO) and IPES (LUSO). 

Figures\ \ref{UPSIPES}b and c clearly show that  for both p$^{++}$ and n$^{++}$ samples, the Fermi level is closer to the onset of the IPES spectrum (LUSO) than to the onset of the UPS spectrum (HOSO) and the estimated uncertainty of the exact position of the HOSO and LUSO is marked by a shaded area. The apparent difference in the edge to edge gap in Fig.\ref{UPSIPES} for the molecular monolayers on p$^{++}$ and n$^{++}$, is within this uncertainty. For p$^{++}$-Si-C$_{16}$H$_{33}$ samples, the energy difference between the Si $E_{\mathrm{F}}$ and the monolayer LUSO ($\sim\unit[3.2]{eV}$), is smaller than that between $E_{\mathrm{F}}$ and the HOSO ($\sim\unit[4]{eV}$). Similarly, for the n$^{++}$-Si-C$_{16}$H$_{33}$ samples, the energy difference between the Si $E_{\mathrm{F}}$ and the monolayer LUSO ($\sim\unit[3]{eV}$) is smaller than that between
the Si $E_{\mathrm{F}}$ and the HOSO ($\sim\unit[4.4]{eV}$).

A prominent feature in Fig.\ \ref{UPSIPES} is that the intrinsic band alignment somewhat varies with  the doping density of Si. In other words, the energy difference between the LUSO (HOSO) and the conduction (valence) band edges of the Si is not constant and shifts with the position of the Fermi level within the Si band gap. This result is surprising, because it is normally assumed that the electronic molecular properties (ionization potential and electron
affinity) are independent of the properties of the substrate. Still, this result is  highly  reproducible experimentally and the same phenomenon was also observed in other hybrid organic-inorganic systems.\cite{avasthi_electronic_2011} Possibly this is related to the above-hypothesised doping-dependent Si-C bond polarization.   Regardless of this result, which is currently under further study, the key experimental finding is that, for both p$^{++}$  and n$^{++}$ Si systems, the Fermi level of the system and the carrier transport level in the Si are significantly closer to the LUSO than to the HOSO.

We now ask if the spectroscopic results presented in Fig.\ \ref{UPSIPES} can efficiently predict the charge transport characteristics of the alkyl chain monolayer?  To answer this, we compare the effective tunneling barrier height that we can deduce from the $J-V$ measurements with the expected  barrier that we find from the IPES spectrum i.e. $|E_{\mathrm{F}}-\mathrm{LUSO}|$ energy difference. To this end we need to take under consideration the Si band bending in the junction at equilibrium (zero bias), rather than the band bending at the Si surface as determined by XPS for the free molecularly- modified Si surface, because that band bending can change after forming a metal (Hg) contact on the molecules. Because the molecular monolayer is very thin ($\unit[4-20]{\AA}$) and supports a large tunneling current, there can be charge transfer between the metal and the Si substrate across the molecules, until electronic equilibrium is established (equalization of the Fermi levels of the Si and the metal). This charge transfer affects the Si band bending, but should not change the alignment between the Si bands (at the surface) and the monolayer HOSO and LUSO, assuming that the molecules do not become charged.

Elsewhere we showed \cite{yaffe_molecular_2009} that, due do the large surface dipole that is introduced by the molecules, depositing Hg on a moderately doped n-Si-C$_{n}$H$_{2n+1}$ sample results in a large band bending in the Si, up to the point where the Si surface is in strong inversion and charge transport in the junction is controlled by minority carrier recombination. For the heavy-doped Si used here, the effect of the band bending on transport is usually much smaller, because the depletion layer is thinner ($\sim \unit[10]{nm}$ compared to $\sim \unit[1]{\upmu m}$ for moderately doped Si) and allows some charge tunneling through it. This type of charge transport mechanism, which is intermediate between thermionic emission and pure tunneling, is known as thermionic field emission (TFE). \cite{rhoderick_monographs_1988,sze_physics_2007}

\begin{figure}
  \includegraphics{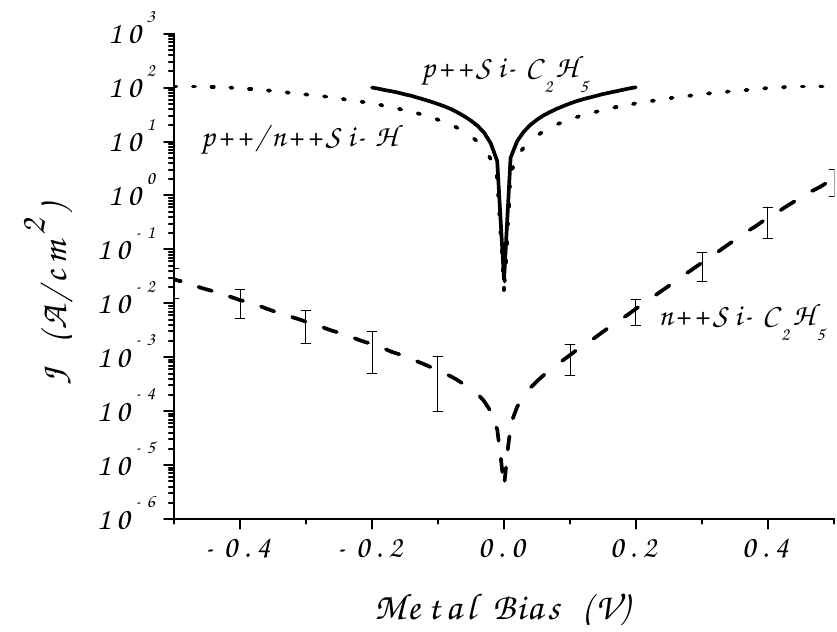}\\
  \caption{$J-V$ measurements of p$^{++}$ and n$^{++}$ Si-C$_{2}$H$_{5}$/Hg junctions (solid and dashed curves, respectively), compared to the J-V behavior of p$^{++}$ and n$^{++}$Si-H/Hg junctions (dotted curve; p$^{++}$ and n$^{++}$ curves are identical). The error bars represent the standard deviation of multiple measurements on 8 different junctions.  }\label{JVshort}
\end{figure}

To isolate the effect of alkyl molecules on the Si band bending from tunneling through the molecules, we performed $J-V$ measurements on a junction with an ultrathin molecular monolayer, C$_{2}$H$_{5}$ (Fig.\ \ref{JVshort}), and compared the results to those obtained with freshly etched, H-terminated Si. Because the C$_{2}$H$_{5}$ monolayer is so thin ($\sim \unit[0.5]{nm}$), it presents a negligible charge transport barrier, compared to the depletion layer in the Si. Combined with CPD evidence that the surface dipole is un-altered by molecular length, the ultra-short alkyls allow quantifying the effect of the depletion layer on net transport.
The $J-V$ characteristics of the Si-H/Hg junction were measured as a control (dotted curve) and found to be Ohmic and identical for p$^{++}$ and n$^{++}$ junctions. The results for Si-C$_{2}$H$_{5}$/Hg are very different with the p$^{++}$ junction exhibiting Ohmic behavior with current density even higher than that of the Si-H samples. The n$^{++}$ junction yields an exponential increase at both bias polarities, a behavior that is typical of charge transport via TFE. \cite{rhoderick_monographs_1988} The high current density of the p$^{++}$ junction is consistent with our assumption that the tunnel barrier presented by the C$_{2}$H$_{5}$ monolayer is negligible, and that the dipole direction works to slightly decrease the Si band bending. The same dipole, on the n$^{++}$-Si surface should, therefore, increase the band bending. Indeed, when we fit the n$^{++}$ results at forward bias (positive bias on the metal) to an analytical expression for TFE developed by Padovani and Stratton, \cite{padovani_field_1966} we find a barrier height of $\unit[0.8]{eV}$.  Because the Si is heavy-doped and the Fermi level is near the CB, this is also roughly the n$^{++}$-Si band bending value at equilibrium.\cite{sze_physics_2007} 
Thus, for p$^{++}$-Si junctions there is negligible internal barrier in the Si, while such a barrier exists for the n$^{++}$-Si junction, where it has a large effect on the overall $J-V$ curve. While this does not necessarily mean that the molecular contribution to transport is overwhelmed by band bending in n$^{++}$-Si, as is the case with moderately doped Si junctions,\cite{yaffe_molecular_2009}  this band bending decreases the current density significantly.  Due to the large internal barrier in the n$^{++}$-Si, we limit ourselves to the molecular length dependent study of the p$^{++}$-Si junction, and then only to the data for positive applied bias on the metal, where according to Fig.\ \ref{NormCond}, current is mostly due to band-to-band tunneling with the density of states available for tunneling approximately constant with applied bias. The reason is that, as we noted earlier, for negative applied bias the dominant charge transport mechanism is ISAT, i.e., transport is controlled by the recombination rate at the Si surface and cannot be described with conventional tunneling models.\cite{freeman_theory_1970}    

Fig.\ \ref{JVlong} presents the $J-V$ behavior of p$^{++}$-Si-C$_{n}$H$_{2n+1}$ /Hg junctions (n=14,16 and 18) (the length dependence of the n$^{++}$-Si-C$_{n}$H$_{2n+1}$ /Hg junctions were published elsewhere \cite{yaffe_hg/molecular_2010}). It is clear that the qualitative behavior of the J-V curve is independent of chain length. It is also clearly seen that, as expected for such junctions, the current density decreases exponentially with the number of methylene units in the alkyl chain. \cite{akkerman_electrical_2008}
This type of behaviour is generally described by the Landauer relation,\cite{adams_charge_2003,heimel_interface_2008,salomon_comparison_2003}  In this model, the conductance of a single channel (G) is given as: \cite{akkerman_electrical_2008,chen_effect_2006,magoga_conductance_1997} 

\begin{equation}\label{1}
G=G_{\mathrm{c}}\exp{(-\beta l)}
\end{equation}

where ‘$l$’ is the tunnel barrier width, which can be taken as the length of the molecule or the thickness of the monolayer, $\beta$ is the length-decay parameter and $G_{\mathrm{c}}$ is the contact conductance. From a $G$ vs. $l$ plot at a given $V_{\mathrm{app}}$, we extract both $\beta$ (slope) and $G_{\mathrm{c}}$ (intercept). We start by discussing the meaning of $G_{\mathrm{c}}$, followed by an interpretation the extracted $\beta$ value. 
The extracted $G_{\mathrm{c}}$ value (at $V_{\mathrm{app}} = \unit[0.1]{V}$) of the p$^{++}$-Si-alkyl/Hg junctions is $2\cdot10^{-4}G_{0}$, where $G_{0}$ is the quantum conductance ($G_{0}\equiv2q^{2}/h=\unit[77.4]{\upmu S} $). The Gc value of n$^{++}$-Si-alkyl/Hg junctions was $4\cdot10^{-7}G_{0}$, \cite{yaffe_hg/molecular_2010} very low compared to typical Au-alkyl MIM junctions. It reflects the large band bending in the Si that was extracted from the $J-V$ behaviour of n$^{++}$ Si-C$_{2}$H$_{5}$/Hg observed in Fig. \ \ref{JVshort}. The $G_{\mathrm{c}}$ found for the p$^{++}$ junction is similar to that reported for an MIM junction with one chemi-contact \cite{akkerman_electrical_2008}, a result that can be rationalized by recalling the near flat band conditions in the p$^{++}$ -Si-C$_{2}$H$_{5}$/Hg junction (Fig.\ \ref{JVshort}). \footnote{While these results suggest that the MIM $G_{\mathrm{c}}$ could serve as a standard value to determine the effect of the band bending on the conduction properties of a molecular MIS junction, that is unlikely to be generally valid. The reason is that $G_{\mathrm{c}}$ should be highly contact dependent and will vary between metals and binding groups, (n.d.).}  
To further verify the correlation between the spectroscopic evidence (Fig.\ \ref{UPSIPES}) and the transport results, we estimate the tunnel barrier ($\phi_{\mathrm{t}}$) from the $J-V$ curves. This barrier should, in principle, reflect the energy difference between the Fermi level of the electrodes and the energy of the nearest allowed state in the interfacial insulator, as detected by UPS and IPES. In addition, we compare the $J-V$ behavior of our p$^{++}$-Si-alkyl/Hg junction to previously reported results on the more common Au-S-alkyl-S-Au junction. To that effect, we use the general WKB approximation for tunneling that describes the relation between $\beta$ and $\phi_{\mathrm{t}}$ as \cite{stratton_volt-current_1962}: 
\begin{equation}\label{2}
\beta_{0}=2\sqrt{2m_{\mathrm{e}}\phi_{\mathrm{t}}/\hbar^2}
\end{equation}
where $\beta_{0}$ is the decay parameter at equilibrium $(V_{\mathrm{app}}\rightarrow0)$, $h$ is Planck's constant, $q$ is the elementary charge and $m_\mathrm{e}$ is the electron mass. This is by no means an accurate model that reflects the complexity of molecular junctions. Nevertheless, as it is the basis for the Simmons model \cite{simmons_generalized_1963} that is widely used for the extraction of $\phi_{\mathrm{t}}$  in molecular junctions, it allows direct comparison with previous results. The difference in dielectric properties between vacuum and insulating materials is commonly accounted for by using an effective mass for the tunneling electron, which is smaller than the free electron mass. Using complex band structure DFT calculations, Tomfohr and Sankey calculated the dispersion relation of the forbidden band gap of an infinite alkyl chain and showed that by using the dispersion relation in vacuum (Eq.\ \ref{2}) with an effective mass of $0.29 m_{\mathrm{e}}$, a lower limit of the barrier is extracted.\cite{tomfohr_complex_2002}  Thus, we use this value of effective mass to extract an estimated value of $\phi_{t}$, which we can compare to previous studies and to the spectroscopic results presented in Fig.\ \ref{UPSIPES}.

The average extracted $\phi_{t}$ in the low bias range ($\unit[0-0.1]{V}$) of the $J-V$ curves in Fig.\ \ref{JVlong} is $\unit[(0.84\pm0.05)]{\AA^{-1}}$. Inserting this value into Eq.\ \ref{2}, along with a $0.29 m_{\mathrm{e}}$ effective mass,\cite{tomfohr_complex_2002}  yields a $\unit[2.3\pm0.3]{eV}$ barrier. Having established that $|E_{\mathrm{F}}-\mathrm{LUSO}|$ dominates charge transport (Fig.\ \ref{NormCond}), we can locate the extracted $\phi_{t}$ value on the IPES spectrum (marked by vertical arrow) in Fig.\ \ref{UPSIPES}. This value is in between the IPES edges and, because it is only the lower limit of the barrier, we conclude that we have reasonable agreement with the LUSO of the alkyl monolayer (edge \#2 in Fig.\ \ref{UPSIPES}). A tunneling barrier of $\sim \unit[2.3]{eV}$ is much smaller than the $|E_{\mathrm{F}}-\mathrm{HOSO}|$ difference of $\unit[4]{eV}$ (Fig.\ \ref{UPSIPES}), decreasing further the possibility of significant HOSO-contribution to transport.

The natural question is how far our observation for LUSO-dominated tunneling in alkyl-Si junctions is applicable to alkyl junctions on any substrate? The answer depends on the position of the molecular levels of the alkyl chains relative to the electrode Fermi level.  UPS/IPES measurements of alkyl thiols on Au \cite{qi_filled_2011-1}  yield a HOSO-LUSO gap of $\unit[7.85]{eV}$ and significantly smaller $|E_{\mathrm{F}}-\mathrm{LUSO}|$ ($\sim \unit[3.35]{eV}$) than $|E_{\mathrm{F}}-\mathrm{HOSO}|$ ($\sim \unit[4.5]{eV}$) (the latter is in good agreement with previously reported $|E_{\mathrm{F}}-\mathrm{HOSO}|$ for alkyl thiols on Au ($\sim \unit[5]{eV}$). \cite{alloway_interface_2003,kera_direct_2001,whelan_benzenethiol_1999}  The HOSO-LUSO gap is similar to the gap measured for alkyl chain monolayers on both Si \cite{salomon_what_2007,thieblemont_electronic_2008}  and GaAs \cite{shpaisman_electrical_2009} with different binding groups (Si-C, Si-O-C and GaAs-PO$_{3}$-C), i.e., \textit{it is not influenced significantly by binding group and substrate type}. The range of the edge-to-edge gap in all examined samples is $\sim \unit[(7.1-7.8)]{eV}$, if the IDIS onset is ignored. These values agree with values reported for polyethylene.\cite{delhalle_electronic_1974-1,fujihira_photoemission_1972,kitani_photoconduction_1980,less_intrinsic_1973}   Thus we conclude that the HOSO-LUSO gap of alkyl monolayers, and their position relative to the vacuum level is rather independent of the contacting electrodes. Therefore, we expect tunneling to be dominated by the $|E_{\mathrm{F}}-\mathrm{LUSO}|$ energy difference for any alkyl junctions, except maybe for electrodes of extremely high work function ($> \sim \unit[6]{eV}$).  

Our conclusion is at odds with earlier studies on molecular transport across alkyl chains, which concluded that transport is dominated by the $|E_{\mathrm{F}}-\mathrm{HOSO}|$ barrier, especially for the extensively studied Au-(alkythiol or alkyldithiol)-Au systems.\cite{beebe_transition_2006,cui_changes_2002,engelkes_length-dependent_2004,huisman_interpretation_2009,li_charge_2008,malen_identifying_2009,paulsson_thermoelectric_2003}   In most cases the evidence for hole tunneling is based on transport across relatively short alkyl chains, up to C12, and is indirect. An exception is the measurement of thermo-power (Seebeck coefficient), where the positive coefficient was taken as direct evidence for hole / HOSO-mediated tunneling. \cite{malen_identifying_2009} Remarkably, though, the thermo-power decreases with increasing molecular length and, when extrapolated, is predicted \textit{to change sign }(i.e., change from HOSO- to LUSO-dominated, hole to electron tunneling) for alkyl chains longer than C12. This length-dependence was rationalized by invoking Metal Induced Gap States (MIGS) as dominating transport, rather than pure molecular states. Such an explanation is very reasonable, because with the short dithiol molecules used in those experiments (Au-SC$_{n}$H$_{2n+1}$S-Au with $n=2,3,4,5,6,8$) MIGS are expected to extend out to several carbon atoms from the contacts. 

In early CP-AFM work on short mono and dithiol alkyl chains the extrapolated zero molecular length resistance increased as the work function of the contacting metal decreased and this was taken to indicate an increasing tunnel barrier.\cite{engelkes_length-dependent_2004} Similar to the thermo-power, this effect can be due to the MIGS, rather than to alkyl-derived energy levels. 

In earlier work with Si as substrate, the type of dominant charge carrier in the semiconductor was suggested to dictate the type of tunneling through (longer) alkyl chains.\cite{selzer_effect_2002}  In later work on a system, similar to the one studied here, the lack of temperature activation was taken as indicating HOSO-mediated /electron tunneling \cite{salomon_temperature-dependent_2008}, an effect that may well be present, but is overwhelmed by the effects that we have shown here. 
Except for the last two reports, we note that the main difference between ours and other experiments is that Si-monolayer/Hg junctions (with a  several $\unit[100]{\upmu m}$ diameter contact, rather than a nm-sized one) make it possible to measure conductance through longer molecules, which allows measuring molecule-, rather than contact-dominated transport by reducing the contribution of MIGS and IDIS interface effects to net transport.

\section{Conclusions} 

To conclude, the analogy to asymmetry in NDR for degenerately-doped MIS of different doping types indicates that for both p$^{++}$- and n$^{++}$-type Si-alkyl chain/Hg junctions charge transport is controlled by $|E_{\mathrm{F}}-\mathrm{LUSO}|$ (rather than by $|E_{\mathrm{F}}-\mathrm{HOSO}|$). Furthermore, by comparing charge transport through p$^{++}$- and n$^{++}$-Si-C2H5/Hg junctions, we showed that there is considerable band bending in the n$^{++}$-Si junction, which has a large effect on the charge transport (even though the Si is heavy doped). The p$^{++}$-Si junction, though, is near flat-band in equilibrium, with comparable transport probability to that of MIM junctions. Using UPS and IPES measurements we elucidated the electronic structure of Si-alkyl system and showed that $|E_{\mathrm{F}}-\mathrm{LUSO}|<|E_{\mathrm{F}}-\mathrm{HOSO}|$, in line with the electron tunneling qualitative argument. The tunneling barrier height estimated from the WKB approximation also fits well with the IPES-derived $|E_{\mathrm{F}}-\mathrm{LUSO}|$ difference.

\section*{Acknowledgements}
We thank the US-Israel Binational Science Foundation (DC, AK), the Israel Science Foundation, ISF, through its Centre of Excellence programs (DC, LK), the Lise Meitner – Minerva Center for Computational Chemistry (LK), the National Science Foundation (DMR-1005892)) and the Princeton MRSEC of the National Science Foundation (DMR- 0819860) (AK), NanoNed, funded by the Dutch Ministry of Economic Affairs (project WSC.6972) (HZ) and the Monroe and Marjorie Burk Fund for Alternative Energy Studies (DC) for partial support. At the Weizmann Inst. this work was made possible in part by the historic generosity of the Harold Perlman family. OY thanks the Azrieli Foundation for an Azrieli Fellowship. DC holds the Sylvia and Rowland Schaefer Chair in Energy Research.

\end{document}